\documentstyle[12pt]{article}
\newcommand{\be}{\begin{equation}}
\newcommand{\ee}{\end{equation}}
\newcommand{\bea}{\begin{eqnarray}}
\newcommand{\eea}{\end{eqnarray}}
\newcommand{\norsl}{\normalsize\sl}
\newcommand{\norsc}{\normalsize\sc}

\begin{document}

\begin{titlepage}

\title{ Positivity Constraints on Photon \\ Structure Functions
}

\author{
\norsc  Ken SASAKI~$^a$\thanks{e-mail address: sasaki@cnb.phys.ynu.ac.jp},~
          Jacques SOFFER~$^b$\thanks{e-mail address: soffer@cpt.univ-mrs.fr}~
and       Tsuneo UEMATSU~$^c$\thanks{e-mail address:
uematsu@phys.h.kyoto-u.ac.jp}
\\
\norsl  $^a$ Dept. of Physics,  Faculty of Engineering, Yokohama National
University \\
\norsl  Yokohama 240-8501, JAPAN \\
\norsl $^b$ Centre de Physique Th{\'e}orique, CNRS, Luminy Case 907,
\\
\norsl        F-13288 Marseille Cedex 9, FRANCE \\
\norsl $^c$ Dept. of Fundamental Sciences, FIHS, Kyoto University \\
\norsl     Kyoto 606-8501, JAPAN \\
}

\date{}
\maketitle

\begin{abstract}
{\normalsize
We investigate the positivity constraints for the structure
functions of both virtual and real photon. From the Cauchy-Schwarz
inequality we derive three positivity conditions for the general
virtual photon case, which reduce, in the real photon case,
to one condition relating the polarized and unpolarized structure functions.

}
\end{abstract}

\begin{picture}(5,2)(-290,-500)
\put(2.3,-65){YNU-HEPTh-01-102}
\put(2.3,-80){CPT-2001/P.4242}
\put(2.3,-95){KUCP-195}
\put(2.3,-110){October 2001}
\end{picture}

\thispagestyle{empty}
\end{titlepage}
\setcounter{page}{1}
\baselineskip 18pt

The photon structure has been studied through the
two-photon processes in e$^+$~e$^-$ collisions as well as the resolved
photon processes in the electron-proton collider. Based on the perturbative
QCD (pQCD), the unpolarized parton distributions in the photon have been 
extracted
from the  measured structure function $F_2^\gamma$ \cite{MK}.
Recently there has been growing interest in the study of polarized
photon structure functions \cite{Barber,SVZ}. Especially the first moment of
the spin-dependent  structure function $g_1^\gamma$ has attracted much
attention in the  literature in connection with its relevance for the axial
anomaly \cite{BASS,ET,NSV,FS,BBS}. The next-to-leading order QCD analysis
of $g_1^\gamma$ has been performed in the literature \cite{SV,SU,GRS}.
There exists a  positivity bound, $|g_1^\gamma|\leq F_1^\gamma$, which
comes out from the  definition of structure functions,
$g_1^\gamma$ and $F_1^\gamma$,  and positive definiteness
of the $s$-channel helicity-nonflip amplitudes.
This bound was closely analyzed recently~\cite{GRS}.

In the case of virtual photon target, there appear eight structure
functions \cite{BCG,BM,CW}, most of which have not been measured yet and,
therefore, unknown. In a situation like this, positivity would play
an important role in constraining these unknown structure functions.
It is well known in the deep inelastic scattering off nucleon
that various bounds have been obtained for the spin-dependent observables and
parton distributions in a nucleon by means of positivity conditions
\cite{JS1}.

In the present paper we investigate the model-independent
constraints for the structure functions of virtual (off-shell) and
real (on-shell) photon target. We obtain  three positivity conditions for
the virtual photon case and one condition for the real photon, the
latter of which relates the polarized and unpolarized structure functions.

Let us consider the virtual photon-photon forward scattering:
$\gamma(q)+\gamma(p)\rightarrow \gamma(q)+\gamma(p)$ illustrated in Fig.1.
The $s$-channel helicity amplitudes are given by
\be
W(ab\vert a'b')=\epsilon^*_\mu(a)\epsilon^*_\rho(b)
W^{\mu\nu\rho\tau}\epsilon_\nu(a')\epsilon_\tau(b')~,
\ee
where $p$ and $q$ are four-momenta of the target and probe photon,
respectively, $\epsilon_\mu (a)$ represents the photon polarization
vector with helicity $a$, and $a, a'=0, \pm1$, and $b, b'=0, \pm1$.
Due to the angular momentum conservation, $W(ab\vert a'b')$ vanishes unless it
satisfies the condition $a-b=a'-b'$. And parity conservation and time reversal
invariance lead to the following properties for
$W(ab\vert a'b')$~\cite{BLS}:
\bea
W(ab\vert a'b')&=&W(-a,-b\vert -a',-b')\qquad {\rm Parity\ conservation~,}
\nonumber  \\
&=&W(a'b'\vert ab)\qquad \qquad\qquad\ \ {\rm Time \ reversal \
invariance~.}
\eea
Thus in total we have eight independent $s$-channel helicity amplitudes,
which we may take as $W(1,1\vert 1,1)$,
$W(1,-1\vert 1,-1)$, $W(1,0\vert 1,0)$, $W(0,1\vert 0,1)$, $W(0,0\vert 0,0)$,
$W(1,1\vert -1,-1)$, $W(1,1\vert 0,0)$, and $W(1,0\vert 0,-1)$.
The first five amplitudes are helicity-nonflip  and the rest are
helicity-flip. It is noted that $s$-channel helicity-nonflip amplitudes
are semi-positive, but not the helicity-flip ones. And
corresponding to these three helicity-flip amplitudes,
we will obtain three non-trivial positivity constraints.

The helicity amplitudes may be expressed in terms of the transition
matrix elements from the state $|a,b\rangle$ of two virtual photons with
helicities $a$ and $b$, to the unobserved state $|X\rangle$ as
\bea
       W(ab|ab)&=&\sum_X |\langle X|a,b\rangle|^2, \nonumber  \\
       W(ab|a'b')&=&{\rm Re}\sum_X \langle X|a,b\rangle^{*} \langle X|a',b'
\rangle
\quad (a\neq a', b\neq b')~.
\eea
Then, a Cauchy-Schwarz inequality~\cite{JS,JT}
\be
\sum_X \Bigl| \langle X\vert a,b\rangle +\alpha
\langle X\vert a',b'\rangle \Bigr|^2 \geq 0~,
\ee
which holds for an arbitrary real number $\alpha$, leads to
a positivity bound for the helicity amplitudes:
$\Bigl|W(a,b\vert a',b')  \Bigr|\leq \sqrt{W(a,b\vert a,b)
W(a',b'\vert a',b')}$~.
Writing down explicitly, we obtain
the following three positivity constraints:
\bea
\Bigl|W(1,1\vert -1,-1)  \Bigr|&\leq& W(1,1\vert 1,1)~,  \label{CS1}\\
\Bigl|W(1,1\vert 0,0)  \Bigr|&\leq& \sqrt{W(1,1\vert 1,1)
W(0,0\vert 0,0)}~,\label{CS2}\\
\Bigl|W(1,0\vert 0,-1)  \Bigr|&\leq& \sqrt{W(1,0\vert 1,0)W(0,1\vert 0,1)}~.
\label{CS3}
\eea
In terms of the eight independent amplitudes introduced by Budnev, Chernyak
and Ginzburg \cite{BCG}, the above three conditions can be rewritten
as
\bea
\Bigl|W_{\rm TT}^\tau \Bigr|&\leq&
\left(W_{\rm TT}+W_{\rm TT}^a\right)~,  \label{BCG1}\\
\Bigl|W_{\rm TS}^\tau +W_{\rm TS}^{\tau a}  \Bigr|&\leq&
\sqrt{(W_{\rm TT}+W_{\rm TT}^a)W_{\rm SS}}~,\label{BCG2}\\
\Bigl| W_{\rm TS}^\tau -W_{\rm TS}^{\tau a} \Bigr|&\leq&
\sqrt{W_{\rm TS}W_{\rm ST}}~,\label{BCG3}
\eea
where T and S refer to the transverse and longitudinal
photon, respectively, and the superscripts "$\tau$" and "$a$" imply the 
relevance
to the helicity-flip  amplitudes and polarized ones, respectively.

For the real photon, $p^2=0$, the number of
independent helicity amplitudes reduces to four. They are
$W(1,1|1,1)$, $W(1,-1|1,-1)$,
$W(0,1|0,1)$, and $W(1,1|-1,-1)$~, which are related to four
structure functions $W_i^{\gamma}$ as follows
\cite{BCG,BM,CW,KS}:
\bea
&&\frac{1}{2}\left[W(1,1|1,1)+W(1,-1|1,-1)\right]
=W_1^\gamma~,  \nonumber\\
&&W(0,1|0,1)
=-W_1^\gamma+\frac{(p\cdot q)^2}{Q^2}W_2^\gamma~,    \nonumber\\
&&\frac{1}{2}W(1,1|-1,-1)=W_3^\gamma~,    \nonumber\\
&&\frac{1}{2}\left[W(1,1|1,1)-W(1,-1|1,-1)\right]
=W_4^\gamma~,
\eea
where the last one is the polarized structure function and
usually denoted by $g_1^\gamma$ with
$W_4^\gamma=\frac{1}{2}g_1^\gamma$~. Also the first one, $W_1^\gamma$,
is often referred to as $F_1^\gamma$ with $W_1^\gamma=\frac{1}{2}F_1^\gamma$~.

For the real photon case we have only one constraint, i.e.,
the first inequality (\ref{CS1}), which is rewritten as
\be
2|W_3^\gamma| \leq (W_1^\gamma +W_4^\gamma)~.\label{realbound}
\ee
It is interesting to recall that the polarized structure function
$W_4^\gamma$ of
the real photon  satisfies a remarkable sum rule~\cite{BASS,ET,NSV,FS,BBS}
\be
\int_0^1 W_4^\gamma (x, Q^2)dx=0~.
\ee
The integral of $|W_3^\gamma|$ is, therefore, bounded from above by the first
moment of $W_1^\gamma$,
\be
\int_0^1 |W_3^\gamma (x, Q^2)| dx\leq \frac{1}{2}\int_0^1 W_1^\gamma (x,
Q^2)dx~.
\ee

Now let us examine whether the inequality (\ref{realbound}) is actually
satisfied or not by the structure functions obtained in the
simple parton model (PM). By evaluating the box (a massive
quark-loop) diagrams with $p^2=0$, ignoring the power
correction of
$m^2/Q^2$ with quark mass $m$, the photon structure
functions have been obtained as follows:
\bea
W_1^{\gamma}(x,Q^2)_{\rm PM}&=&\frac{\alpha}{2\pi}\delta_{\gamma}
\biggl\{\Bigl[ x^2+(1-x)^2  \Bigr]{\rm ln} \biggl(\frac{Q^2}{m^2}
\frac{1-x}{x}  \biggr)
-1+4x(1-x)  \biggr\}~,    \nonumber \\
W_3^{\gamma}(x,Q^2)_{\rm PM}&=&\frac{\alpha}{2\pi}\delta_{\gamma}
(-x^2)~,  \label{parton} \\
W_4^{\gamma}(x,Q^2)_{\rm PM}&=&\frac{\alpha}{2\pi}\delta_{\gamma}
\biggl\{(2x-1){\rm ln} \biggl(\frac{Q^2}{m^2} \frac{1-x}{x}  \biggr)
+3-4x  \biggr\}~,\nonumber
\eea
where $x=Q^2/(2p\cdot q)$, $\alpha=e^2/4\pi$, the QED coupling constant, and
$\displaystyle{\delta_{\gamma}=3\sum_{i=1}^{N_f} e_i^4}$,
with $N_f$, the number of the active flavors.
Using these expressions, we examine the constraint (\ref{realbound})
numerically and find that it is satisfied almost all allowed region of $x$
except near  the limit $x\rightarrow x_{\rm max}=1/(1+4m^2/Q^2)$.
However, the violation of the inequality near $x_{\rm max}$ is an artifact,
since the limiting procedures of $Q^2 \rightarrow \infty$ and
$x\rightarrow x_{\rm max}$ are not exchangeable. In fact, the exact
PM calculation of $W_i^\gamma$'s  with $Q^2$ kept finite gives
\bea
W_1^{\gamma}\vert_{\rm PM}&=&\frac{\alpha}{2\pi}\delta_{\gamma}
\biggl\{\biggl(
    {\rm ln} \frac{1+\beta}{1-\beta}\biggr)\Bigl[x^2+(1-x)^2
-8x^2\frac{m^4}{Q^4}  -4(x^2-x)\frac{m^2}{Q^2} \Bigr] \nonumber  \\
&& \qquad \qquad +\beta \Bigl[4x(1-x)-1 +4(x^2-x)\frac{m^2}{Q^2}\Bigr]
\biggr\}~,  \nonumber  \\
W_3^{\gamma}\vert_{\rm PM}&=&-\frac{\alpha}{2\pi}\delta_{\gamma}
\biggl\{\biggl(
    {\rm ln} \frac{1+\beta}{1-\beta}\biggr)\Bigl[ 4x^2\frac{m^4}{Q^4}
    +4x^2\frac{m^2}{Q^2} \Bigr]  +\beta \Bigl[x^2 +2(x-x^2)\frac{m^2}{Q^2}
\Bigr]\biggr\}~,\nonumber\\
W_4^{\gamma}\vert_{\rm PM}&=&\frac{\alpha}{2\pi}\delta_{\gamma}\biggl\{\biggl(
    {\rm ln} \frac{1+\beta}{1-\beta}\biggr)(2x-1)+\beta \Bigl[-4x+3\Bigr]
\biggr\}~,\label{exactPM}
\eea
where  $\beta=\sqrt{1-\frac{4m^2x}{Q^2(1-x)}}$. The above results
are in accord with the cross sections for the
$\gamma\gamma \rightarrow e^+e^- (\mu^+\mu^-)$ process obtained
by Budnev et al.\cite {BGMS}. Also the expression of
$W_4^{\gamma}\vert_{\rm PM}$ is consistent with the result of Ref.\cite{GRI},
where polarized gluon structure functions were considered.
It is noted that since $\beta\rightarrow 0$ for $x
\rightarrow x_{\rm max}$, all $W_1^{\gamma}\vert_{\rm PM}$,
$W_3^{\gamma}\vert_{\rm PM}$, and
$W_4^{\gamma}\vert_{\rm PM}$ vanish at $x=x_{\rm max}$~.
Using these exact PM results in (\ref{exactPM}),
we find numerically that the inequality (\ref{realbound}) is
indeed satisfied for all allowed  region of $x$~.
Moreover, once expressed as functions of $x$ and $\beta$,
the helicity-nonflip amplitudes $W(1,1|1,1)|_{\rm PM}$ and
$W(1,-1|1,-1)|_{\rm PM}$ are easily shown to be non-negative
for  $0\leq\beta<1$, and  $0\leq x<1$, as they should be.
On the other hand, the helicity-flip amplitude $W(1,1|-1,-1)|_{\rm PM}$
turns out
to be negative.

As stated earlier, in the case of virtual photon,
$p^2=-P^2\neq 0$, there appear
eight structure functions (four of them are new) and we have derived three
positivity  constraints on these functions. But up to now little
attention has been paid  to the virtual photon case and, therefore,
we have slight knowledge of the new photon structure functions. In this
situation it is worthwhile to investigate these new structure functions in
the simple PM and examine that the three positivity constraints
(\ref{BCG1})-(\ref{BCG3}) actually hold~\cite{SSU}.

Especially, in the kinematical region, $\Lambda^2\ll P^2 \ll Q^2$, where
the mass squared of the target photon ($P^2$) is much bigger than the
QCD scale parameter ($\Lambda^2$), some of the photon
structure functions are predictable in pQCD entirely up to the
next-leading-order
(NLO), since the hadronic component on the photon can also be dealt with
perturbatively. Following this strategy, the  virtual photon structure
functions, unpolarized $F_2^\gamma (x,Q^2,P^2)$ and $F_L^\gamma
(x,Q^2,P^2)$~\cite{UW} and  polarized  $g_1^\gamma (x,Q^2,P^2)$~\cite{SU},
were studied up to the NLO. Since $F_1^\gamma \equiv
(F_2^\gamma-F_L^\gamma)/x=2W_1^\gamma$ and $g_1^\gamma=2W_4^\gamma$, it
is also interesting to see if the inequality (\ref{realbound}) is satisfied
by the pQCD results for the above kinematical region~\cite{SSU}.
The virtual photon structure function $W_3^\gamma(x,Q^2,P^2)$ is expected to be
given by the same expression as the PM result (\ref{parton})
up to  ${\cal O} (1/\ln (Q^2/\Lambda^2))$, since there exist
no  twist-2 quark operators contributing to $W_3^\gamma$~\cite{KS,MANO}.

So far we have only considered the constraints on the
structure functions. Now our argument can be extended
to the quark contents of the photon, for which we can also
write down inequalities involving various distributions.
Following Ref. \cite{JS}, let us define the helicity amplitudes given by
\bea
{\cal W}(ab|ab)&\equiv& \sum_X\langle \gamma_b|O^\dagger|q_a,X\rangle
\langle X,q_a|O|\gamma_b\rangle~,\\
{\cal W}(ab|a'b')&\equiv& {\rm Re}\sum_X\langle \gamma_b|O^\dagger|q_a,X\rangle
\langle X,q_{a'}|O|\gamma_{b'}\rangle \quad (a\neq a',\ b\neq b')~,
\eea
where all the suffices, $a$, $b$, $a'$ and $b'$,
refer to the helicities of the quarks and virtual photons, and $O$'s
denote bilinear quark operators.
One also has to sum over all intermediate states $X$.
Then we can derive, in a similar fashion based on the Cauchy-Schwarz 
inequality,
\be
|{\cal W}(ab|a'b')|\leq \sqrt{{\cal W}(ab|ab){\cal W}(a'b'|a'b')}~.\label{CSQ}
\ee

In our present case, the above helicity amplitudes
become nothing but the following quark distributions:
\bea
q^{\pm}_{\gamma}&=& \sum_{X}\langle \gamma_{+}|O^\dagger
|q_{\pm}, X\rangle
\langle X, q_{\pm}|O|\gamma_{+}\rangle~,\\
q^{0}_{\gamma}&=& \sum_{X}\langle \gamma_{0}|O^\dagger
|q_{+}, X\rangle
\langle X, q_{+}|O|\gamma_{0}\rangle~,\\
h^{q}_{\gamma}&=& {\rm Re}\sum_{X}\langle \gamma_{+}|O^\dagger
|q_{+}, X\rangle
\langle X, q_{-}|O|\gamma_{0}\rangle~,
\eea
where $q_\gamma^{\pm}$ ($q_\gamma^0$) denotes
the longitudinally (transversely) polarized quark distribution
inside the photon, and $h_\gamma^q$ is the (chirality-odd)
transversity distribution, the photon analog of $h_1^q$ for
the nucleon case.
Here we note that the photon structure function $F_1^\gamma$
($g_1^\gamma$) can be expressed as a sum over the active quark (or anti-quark)
distributions
$q_{\gamma}$ ($\Delta q_{\gamma}$), with
$q_{\gamma} = q^+_{\gamma} + q^-_{\gamma}$ and $\Delta
q_{\gamma} = q^+_{\gamma} - q^-_{\gamma}$.
Now taking $a=1/2,\ b=1,\ a'=-1/2,\ b'=0$ in (\ref{CSQ}) we get
\be
|h^q_\gamma|\leq \sqrt{q^+_\gamma q^0_\gamma}~.
\ee
Hence we have the following positivity condition for
the transversity distribution $h^q_\gamma$,
\be
|h^q_{\gamma}| \leq \sqrt {(\frac {q_{\gamma} + \Delta q_{\gamma}}{2})
\cdot q^{0}_{\gamma}}~.
\ee
This is an extension of the inequality obtained for the nucleon case
\cite{JS1,JS,JT}\footnote{After this paper was completed, we were informed that Eq.(24) 
   coincide with a result obtained for distribution functions of
   spin-one hadrons, see A. Bacchetta and P.J. Mulders, Phys. Lett.
   B518 (2001) 85.}.
The transversity distribution $h^q_\gamma$ of the photon
could be measured by the semi-inclusive process in the two-photon
reactions provided by the future polarized e$^+$e$^-$ collision experiments.

In summary we have investigated the model-independent
positivity constraints for the
photon structure functions which could be studied in future experiments.
We also discussed a positivity bound for the quark distributions relevant
for the spin-dependent semi-inclusive process in two-photon reactions.
We expect these bounds would provide uselful constraints for studying the
yet unknown polarized and unpolarized photon structures.

\newpage

\vspace{0.5cm}
\leftline{\large\bf Acknowledgement}
\vspace{0.5cm}

We thank Werner Vogelsang and Hideshi Baba for useful discussions.
This work is partially supported by the Grant-in-Aid
from the Japan Ministry of Education and Science,
No.(C)(2)-12640266.


\newpage


\newpage
\vspace{3cm}
\noindent
{\large Figure Captions}
\baselineskip 16pt

\begin{description}

\item[Fig.1] \quad

Virtual photon-photon forward scattering with momenta $q(p)$
and helicities $a(b)$ and $a'(b')$.

\end{description}

\pagestyle{empty}
\input epsf.sty
\begin{figure}
\centerline{
\epsfxsize=11cm
\epsfbox{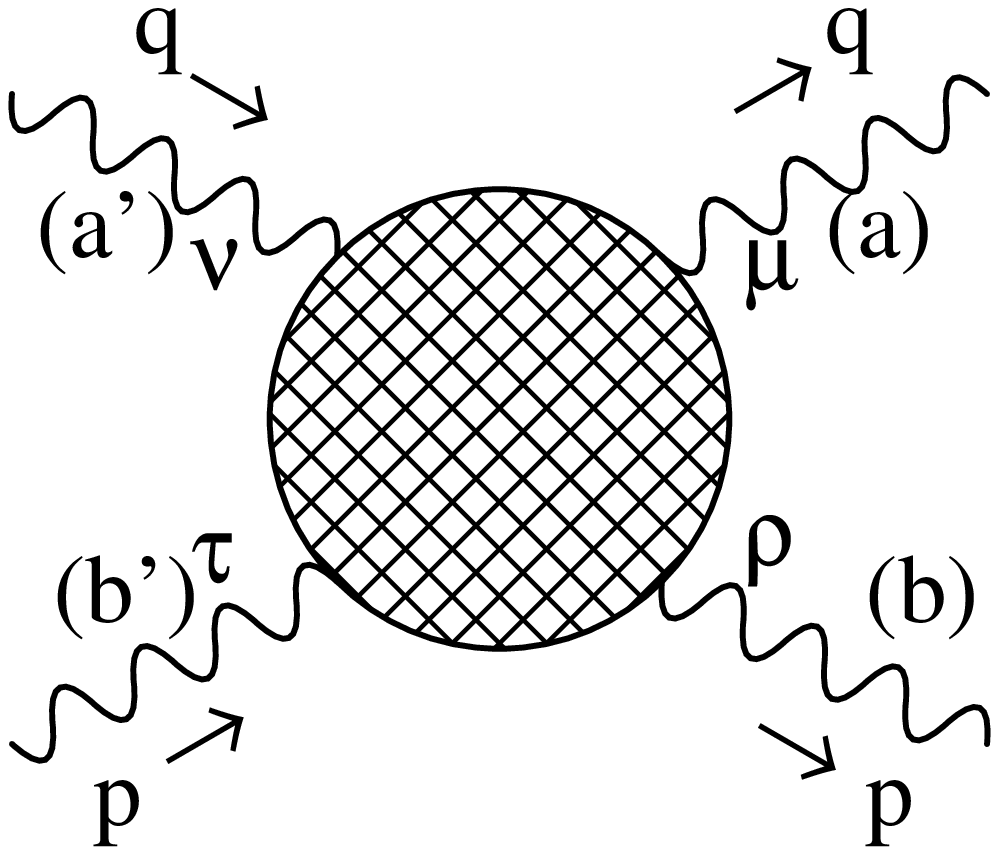}}
\end{figure}


\begin{thebibliography}{99}

\bibitem{MK}
M.~Krawczyk, Talk at PHOTON 2000, Ambleside, England, 26-31 August 2000;
hep-ph/0012179, to be published in the AIP Conference Series CP571 (2001),
ed. A.~J.~Finch, and references therein.

\bibitem{Barber}
        D.~Barber, in proceedings of the \lq\lq Zeuthen Workshop on the
        Prospects of Spin Physics at HERA\rq\rq, DESY  95-200, p.76,
        eds. J.~Bl\"{u}mlein and W.~D.~Nowak.

\bibitem{SVZ}
        M.~Stratmann and W.~Vogelsang, {\sl Z.~Phys.} {\bf C74} (1997) 641.

\bibitem{BASS}
        S.~D.~Bass, {\sl Int.~J.~Mod.~Phys.} {\bf A7} (1992) 6039.

\bibitem{ET}
        A.~V.~Efremov and O.~V.~Teryaev,  {\sl Phys.~Lett.} {\bf B240}
        (1990) 200.

\bibitem{NSV}
        S.~Narison, G.~M.~Shore and G.~Veneziano,
         {\sl Nucl.~Phys.} {\bf B391} (1993) 69; \\
        G.~M.~Shore and G.~Veneziano, {\sl Mod.~Phys.~Lett.}
         {\bf A8} (1993) 373;\\
        G.~M.~Shore and G.~Veneziano, {\sl Nucl.~Phys.} {\bf B381} (1992) 23.

\bibitem{FS}
        A.~Freund and L.~M.~Sehgal, {\sl Phys.~Lett.} {\bf B341} (1994) 90.

\bibitem{BBS}
        S.~D.~Bass, S.~J.~Brodsky and I.~Schmidt,
         {\sl Phys.~Lett.} {\bf B437} (1998) 424.

\bibitem{SV}
        M.~Stratmann and W.~Vogelsang, {\sl Phys.~Lett.} {\bf B386} (1996) 370.

\bibitem{SU}
        K.~Sasaki and T.~Uematsu, {\sl Phys.~Rev.} {\bf D59} (1999) 114011;
        {\sl Phys. Lett.} {\bf B473} (2000) 309; {\sl Eur.~Phys.~J.}
        {\bf C20} (2001) 283.

\bibitem{GRS}
        M.~Gl{\"u}ck, E.~Reya and C.~Sieg, {\sl Phys.~Lett.} {\bf B503} (2001)
        285; {\sl Eur.~Phys.~J.}  {\bf C20} (2001) 271.

\bibitem{BCG}
V.~M.~Budnev, V.~L.~Chernyak and I.~F.~Ginzburg, {\sl Nucl.~Phys.}
{\bf B34} (1971) 470.

\bibitem{BM}
        R.~W.~Brown and I.~J.~Muzinich, {\sl Phys.~Rev.} {\bf D4} (1971) 1496.

\bibitem{CW}
        C.~E.~Carlson and W.~K.~Tung, {\sl Phys.~Rev.} {\bf D4} (1971) 2873.

\bibitem{JS1}
        J.~Soffer, in: K.~Hatanaka, T.~Nakano, K.~Imai, H.~Ejiri (Eds.),
        Proc. of SPIN 2000  (Osaka, 2000), AIP Conference Series CP570, p.461~.

\bibitem{BLS}
        C.~Bourrely, E.~Leader, J.~Soffer, {\sl Phys.~Rep.} {\bf 59} (1980) 95.


\bibitem{JS}
        J.~Soffer, {\sl Phys.~Rev.~Lett.} {\bf 74} (1995) 1292.

\bibitem{JT}
        J.~Soffer and O.~V.~Teryaev, {\sl Phys.~Lett.} {\bf B419} (1998) 400;
            {\sl Phys.~Lett.} {\bf B490} (2000) 106.


\bibitem{KS}
          K.~Sasaki, {\sl Phys.~Rev.} {\bf D22} (1980) 2143;
            {\sl Prog.~Theor.~Phys.~Suppl.} {\bf 77} (1983) 197.
            
\bibitem{BGMS}
          V.M. Budnev, I.F. Ginzburg, G.V. Meledin and V.G. Serbo, {\sl Phys.~Rep.} 
          {\bf 15} (1975) 181.            


\bibitem{GRI}
        A.~Gabrieli and G.~Ridolfi, {\sl Phys.~Lett.} {\bf B417} (1998) 369.
        

\bibitem{SSU}
        K.~Sasaki, J.~Soffer, T.~Uematsu, in preparation.
        

\bibitem{UW}
          T.~Uematsu, T.~F.~Walsh, {\sl Phys.~Lett.} {\bf B101} (1981) 263;
            {\sl Nucl.~Phys.} {\bf B199} (1982) 93.


\bibitem{MANO}
       A.~V.~Manohar, {\sl Phys.~Lett.} {\bf B219} (1989) 357.


\end{thebibliography}
\end{document}